\documentclass{article}
\usepackage{sao2}
\usepackage{psfig}
\issue{2003, 57, 5-163}
\begin{document}
\markboth{MITRONOVA ET AL.}{THE 2MASS-SELECTED FLAT GALAXY CATALOG}
\title{The 2MASS-selected Flat Galaxy Catalog}
\author{S.N. Mitronova \inst{a,d}
\and I.D. Karachentsev \inst{a} \and V.E. Karachentseva \inst{b}
\and T.H. Jarrett  \inst{c}
\and Yu.N. Kudrya \inst{b}}
\institute{Special  Astrophysical Observatory of the Russian Academy
of Sciences, Nizhnij Arkhyz, KChR, 369167, Russia
\and  Astronomical Observatory of Kiev University, Observatorna 3,
   Kiev 04053, Ukraine
\and Infrared Processing
and Analysis Center, Mail Stop 100-22, California Institute of
Technology, Jet Propulsion\\ Laboratory, Pasadena, CA 91125
\and Isaac Newton Institute of Chile, SAO Branch, Russia}
%\date{today}
\maketitle
{\bf Abstract.}
An all-sky catalog of 18020 disc-like galaxies is presented.
The galaxies are selected from the Extended Source Catalog of the Two Micron
All-Sky Survey (XSC 2MASS) basing on their 2MASS axial ratio $a/b\geq3$.
The Catalog contains data on magnitudes of a galaxy in the J, H, K$_s$ bands,
its axial ratio, positional angle, index of luminosity concentration,
as well as identification of the galaxy with the LEDA and the NED databases.
Unlike the available optical catalogs, the new 2MFGC catalog seems to be more
suitable to study cosmic streaming on a scale of $z\la0.1$. The dipole
moment of distribution of the bright (K$<11^m$) 2MFGC objects ($l=227\degr,
b=41\degr$ or $SGL=90\degr, SGB=-43\degr$) lies within statistical errors
($\pm15\degr$) in the direction of the IRAS dipole and the optical RFGC dipole.

\newline

%{\bf Аннотация.}
%Представлен каталог 18020 дискообразных галактик 2MFGC, охватывающий
%все небо. Галактики были отобраны из каталога протяженных источников
%2MASS-обзора (XSC 2MASS) по критерию отношения их осей в 2MASS, $a/b\geq3$.
%Каталог содержит данные о J, H, K$_s$ звездных величинах, отношениях осей,
%позиционных углах, индексе концентрации света, а также отождествления
%2MFGC-галактик в базах данных LEDA и NED. В отличие от существующих
%оптических каталогов, новый (2MFGC) каталог представляется более подходящим
%для изучения космических течений на шкале $z\la0.1$. Дипольный момент
%распределения ярких (K$<11^m$) 2MFGC объектов ($l=227\degr,
%b=41\degr$, или $SGL=90\degr, SGB=-43\degr$) расположен в пределах
%статистических ошибок ($\pm15\degr$) в направлении IRAS-диполя и
%оптического RFGC-диполя.

\vspace*{-0.3cm}
\section{Preface}
The catalog of flat spiral edge-on galaxies, FGC (Karachentsev et al. 1993),
was created as a result of total examination of blue and red prints of the
sky surveys POSS-I and ESO/SERC. The catalog is a rather specific sample of
4455 spiral galaxies with a limiting  blue diameter $a_{lim}=0\farcm6$
and an axial ratio $a/b\ge7$. It consists of two parts: FGC (Flat Galaxy
Catalogue) itself, which covers a region $\delta > -20\degr$, and its
southern extension FGCE (Flat Galaxy Catalogue Extension) with
$\delta < -20\degr$. Since the limiting angular diameter in the catalog
$a_{lim}=0\farcm6$ is smaller than $a_{lim}=1\farcm0$ in the catalogs UGC
(Nilson 1973) and ESO/Uppsala (Lauberts 1982), then more than half (56\,\%)
of flat galaxies entering into it have been cataloged for the first time.

The axial ratio $a/b$ presented in FGC is actually $R_{25}=D_{25}/d_{25}$
in RC3 (de Vaucouleurs et al. 1991) and virtually characterizes the
relationship between the disc and spherical components in a spiral galaxy
viewed edge-on. The condition $a/b\geq 7$ excludes the majority of early-type
spiral galaxies. That is why the average morphological type of the
FGC(E) galaxies is between Sc and Scd, and above 75\% of the FGC galaxies are
spirals of type  Sc and later.

As has been shown by Karachentsev et al. (1997), the mean axial ratios of the
blue images of the galaxies exceed the corresponding axial ratios on the red
prints of both parts of the Catalog.
 This can readily be explained by that the spherical
component of galaxies is redder. When passing to the infrared range, this
feature is intensified, which will be discussed in detail in the next section.

The new improved and complemented version of the catalog,  RFGC (Revised Flat
Galaxy Catalogue --- Karachentsev et al. 1999), involves  4236 flat galaxies.
The principal changes made in RFGC in comparison with FGC are as follows.

\begin{itemize}
\item On the Digital Sky Survey (DSS) the coordinates of all the objects were
measured anew with a higher ($\simeq3\arcsec$) accuracy.
\item The diameters measured on J, R films of the survey ESO/SERC were reduced
to the system of diameters POSS-I (which approximately corresponds to the
system $a_{25}$) based on the results from Kudrya et al. (1997a). This
eliminated the difference in photometric depth between the two
parts of the catalog, FGC and FGCE. About 200 galaxies having reduced
diameters $a<0\farcm6$ were not included in RFGC.
\item For the  RFGC galaxies total apparent magnitudes $B_t$ consistent to an
accuracy of  $\sim0\fm25$ with $B_t$ magnitudes of RC3 (de
Vaucouleurs et al. 1991) were calculated. Angular diameters, surface
brightnesses and other parameters were used in the calculations (Kudrya et al.
1997b).
\end{itemize}

For the present time RFGC is the deepest morphologically homogeneous sample
of the field spiral galaxies. It is mainly intended for studying large-scale
flows of galaxies, properties of the structures in discs etc.
(Karachentsev 1989; Karachentsev et al. 2000; Bizyaev \& Mitronova
2002; Zasov et al. 2002).

\vspace*{-0.3cm}
\section{RFGC galaxies in the survey  2MASS}
The Two Micron All-Sky Survey
(2MASS)  was carried out in three
infrared bands: J (1.11--1.36$\mu$), H (1.50--1.80$\mu$) and
K$_s$ (2.00--2.32 $\mu$)  with two 1.3\,m telescopes in Arizona and Chile,
respectively (Skrutskie et al. 1997). Approximately 3 million galaxies brighter
 K$_s = 14.5^m$ were detected in the  2MASS survey (Jarrett 2000). About 1.65
million galaxies with K$_s<14^m$  and angular dimensions above 7$\arcsec$
were entered into the 2MASS Extended Sources Catalog XSC (Cutri \& Skrutskie
1998). Selection of extended sources and their photometry were performed with
using the standard set of algorithms (Jarrett et al. 2000)
 and the photometric calibration was described in the paper by
 Nikolaev et al. (2000).

Jarrett (2000) notes that the 2MASS survey has weak sensitivity to late-type
galaxies, especially of low surface brightness. The cause lies in high
brightness of the night sky in the near-IR range and short exposures
($\sim$8\,s/object). For this reason the periphery of the discs of spiral
galaxies is generally unseen on isophotes fainter than K = $20^m/\sq\arcsec$.
This conclusion was corroborated and extended for the sample of
 RFGC galaxies (Karachentsev et al. 2002). Mutual identification of the
objects from RFGC and 2MASS  XSC showed that
  2996 out of 4236 (i.e. 71\%) RFGC galaxies were identified in  XSC.
The portion of  RFGC galaxies in
 2MASS decreases
\begin{itemize}
\item when passing from bright galaxies to faint ones (100\% of detection
at $B_t \leq 14\fm5$ and 38\% at $B_t =17\fm5$),
\item from high surface brightness galaxies to galaxies with low surface
brightness
(89\% of detection in RFGC with surface brightness class  SB=I and  57\%
with class SB=IV),
\item from early morphological types to later ones (100\% of detection for
type Sab and 22\% for Sm).
\end{itemize}

A comparison of infrared and optical diameters for nearby large galaxies has
shown that 2MASS diameters are typically $\sim 50$ to 70\,\% relative to
optical measurements, from late to early-type disc spirals, respectively
(see Fig.\,24 of Jarrett et al. 2003). Likewise, 2MASS diameters of
RFGC galaxies are,
on average, half as much as standard optical
diameters.
In the RFGC galaxies the blue axial ratios  $a/b$ cover an interval of
[7--21] with a median of
8.6, while the respective infrared $a/b$  occupy a range of [1--10] with
a median of 4.1.
The comparison  of infrared and optical characteristics made for the RFGC
galaxies resulted in a knowledge what  flat spiral
galaxies look like in the infrared region.

 At the next stage, we applied the data of 2MASS
photometry of the  RFGC galaxies to calculating parameters of their bulk motion.
For the RFGC galaxies with known hydrogen line widths,
Tully--Fischer relationships (TF) in B, I, J, H, K bands
were derived (Karachentsev et al. 2002).
Then, using the J, H, K Tully--Fischer relationships, Kudrya et al.
 (2003) obtained
 the absolute value of the bulk velocity and  the apex parameters  for a
sample of 971 RFGC galaxies with radial velocities $V_{3K} \leq 18000$\,km\,s$^{-1}$:
$V= 199\pm61$\,km\,s$^{-1}$, $l = 301\degr\pm18\degr$, $b=-2\degr\pm15\degr$.
 This is in a good
agreement with our previous estimates obtained when using the TF relationship
for optical diameters (Karachentsev et al. 2000) and also results of other
authors.

Thus, we proved the homogeneity over the sky and the photometric
homogeneity of the catalog 2MASS -- XSC  by the
example of the flat galaxies of the catalog
RFGC.
This induced us to make a more representative and deep catalog of ``flattened''
galaxies using the data of the infrared catalog 2MASS - XSC.
%\end{document}
\section{Catalog 2MFGC}
\subsection{ Criterion of selection and catalog characteristic features}
Among numerous photometric parameters of galaxies included in
XSC (Jarrett et al. 2000) we selected, as reference for the new catalog, a
parameter that characterized the ratio of the minor to major axis. It can
be expressed as the axial ratio of a galaxy image in each band
 $J, H, K$  ($jba, hba, kba$) and also as the axial ratio for the combined
$J+H+K$ image, $sba$ (``super'' coadd).

From the XSC objects were preliminary selected those  whose axial ratio values
even in one band or
$sba$ were smaller than or equal to 0.4,
which significantly  reduced the possibility of
missing  a disc-like galaxy in the list because of features of
photometric processing. The selected objects are present in two lists: the main
one (N = 20631), which included galaxies reliably classified in XSC, and the
additional one (N = 2199), in which binary and multiple stars, as well as
artifacts, could be involved along with galaxies.

Based on the data of Section\,2, galaxies with axial ratio in the blue region
of the spectrum $a/b\simeq6$, will have $a/b=2.86$ or $b/a=0.35$ in 2MASS.
Allowing for the step-type behavior (0.02) of the $a/b$ value in the XSC,
 we adopted the upper
limit of the ratio of the minor-to-major axis to be equal to 0.34
as a  criterion of selection of objects for our Catalog.

The following selection criteria were further applied to the objects of these
lists:

1) $sba\leq0.34$,

2) $sba>0.34$, but the axial ratio in each of the bands
$jba, hba, kba$ was less than or equal to 0.34.

Objects for which even one value of
$jba, hba, kba$ equaled 1 and objects without the estimate of the axial ratio
were excluded from consideration. The limit was not specified either by the
magnitude or by the angular diameter
\footnote{Out of 2199  objects of the additional list the criterion was applied
only to 508 objects identified in LEDA. Only 131 galaxies of them satisfied
criteria 1 and 2.}

Finally, 18215 objects from the main list complied with criteria 1 and 2,
and 16132 of them were identified in the databases LEDA and/or NED. Here,
the radius of the
region of identification was chosen to be 1$\arcmin$ for NED, and
 0.5$\arcmin$ for LEDA.

When two or more objects fell within the region of identification, their
characteristics were visually estimated from the images of the digital sky
surveys POSS-II, ESO/SERC and IPAC 2MASS. The Catalog included the objects
whose characteristics (position angle, axial ratio, brightness, size) fitted
best those presented in XSC.

The same procedure was carried out for single  XSC objects which were
identified with low (worse than 0.4$\arcmin$) accuracy in LEDA and NED.

Note that when creating  any catalog, errors of the Ist and IInd type are
bound to arise: I --- ``true'' objects are not included, II --- ``spurious''
objects are included. As  applied to our Catalog, these errors manifest
themself as follows.

I. In the course of photometric processing the peripheral regions are
inadequately allowed for in galaxies with low surface brightness and the
object  formally ``becomes round''. The selection criterion cuts off a
certain part of really flattened galaxies thereby strengthening the
selection effect already present in the catalog XSC. Besides, the procedure
of measuring galaxies with the greatest angular sizes proved to be
non-optimal for the determination of their parameters. These galaxies are
supposed to be measured in another manner, and they will constitute a
separate 2MASS catalog. It will also  include flattened galaxies that failed
to fall within our Catalog.

II. The photometric procedure fails to distinguish between a single object and
cases where a galaxy is in contact with another galaxy or a ster.
 And, vice versa, a bright detail
can be measured separately from the main body of the  galaxy.

To diminish the influence of errors of the Ist and IInd type at different
stages of compiling the catalog 2MFGC (2MASS Flat Galaxy Catalog)
several thousand of galaxy images were revised  on the frames of 2MASS.
As a result, we included 18020
galaxies in the catalog 2MFGC.

The new catalog 2MFGC comprises 2371  RFGC galaxies, which accounts for
56\% of all
RFGC galaxies and a total of 13\% of the whole number of ``flattened''  2MASS
galaxies.

 In Section\,2 we presented 2996 RFGC galaxies (71\%  of the total number
 of 4236)
identified with
XSC.  As an additional check showed, 18\% out of
2996 galaxies have the axial ratio $sba\geq0.4$. Thus, the ``loss'' of
15\% of really flat galaxies may be due to the application of the axial ratio
as the criterion of  selection of galaxies
for the catalog
2MFGC.

Some typical images of the galaxies from the catalog 2MFGC are shown in
 Fig.\,1.

{\em 2MFGC~2654$\equiv$2dFGRSS320Z078}. $V_h= 31735$\,km/s,
mag=$17\fm77$ (NED), $K = 12\fm89$. An example of a distant galaxy.

{\em 2MFGC~6265$\equiv$ESO~059-022}. $Sbc, V_h=10639$ km/s,
$B_t = 15\fm29$, $K=10\fm61$. A typical example of a moderately distant galaxy.

{\em 2MFGC~6271$\equiv$NGC~2470.} $Sab, V_h=4054$\,km/s, $B_t = 13\fm64$,
$K = 9\fm29$. An example of an early-type spiral galaxy with reliably measured
characteristics.

{\em 2MFGC~6472.} Unidentified in the optical range. $K = 10\fm21$. An example
of a galaxy of very low surface brightness, possibly, a nearby object.

{\em 2MFGC~6562$\equiv$RFGC~1355}.  $Scd$, $B_t=16\fm5$, $K = 13\fm86$. A typical
flat galaxy, seen at the  2MASS limit.

{\em 2MFGC~7000$\equiv$PGC~3085218}. $B_t=21\fm77\pm1\fm0$, $K=11\fm34$.
In the optical range is seen at the limit.

{\em 2MFGC~10056$\equiv$RFGC~23673}. $Sb, V_h=5285$ km/s, $B_t=14\fm5$,
$K = 9\fm65$. A typically flat galaxy, equally well visible in the optical and
infrared ranges.

{\em 2MFGC~13442$\equiv$ESO~179-013}. $SB(s)dm. V_h=834$ km/s,
 $B_t=14\fm88\pm2\fm3$,
$K = 11\fm26$. A nearby galaxy of low surface brightness, is seen through
 the Milky Way.

In the columns of the Catalog we present the following data:

column 1 : 2MFGC galaxy number;

column 2 : Right Ascension and Declination for the epoch
  J2000.0;

column 3 : $K$-band fiducial elliptical Kron radius in arcsecond;

column 4--6 : $J, H, K$-band fiducial elliptical Kron magnitudes;

column 7 :   axis ratio (b/a) for the $J + H + K$ combined image
   (``super'' coadd);

column 8 :  axis ratio (b/a) averaged by the individual $J, H$, and $K$-band
  images;

column 9 :  positional angle in degrees for the ``super'' coadd
  image measured towards NE (+) and NW (--);

column 10 : $ J$-band concentration index (3/4 vs 1/4 light radii);

column 11 :  galaxy number in the LEDA database;

column 12 :  distance from the 2MASS center to the LEDA object
   in arcmin;

column 13 :  galaxy name in the NED database;

column 14 :  distance from the 2MASS center to the NED object center
  in arcmin.

Note that 2877 2MFGC  galaxies, i.e. 16\%, are not presented in the known
catalogs.

\vspace*{-0.3cm}
\subsection{Statistics of the main parameters}

Fig.\,2 presents integral distributions of the number of 2MFGC
galaxies by apparent magnitudes. The straight line represents homogeneous
distribution. As can be seen, the incompleteness of the sample for flattened
galaxies becomes noticeable at $K\simeq13^m$, that is, at 100\% completeness
of the catalog  XSC up to $K\simeq1^m$ (Jarrett 2000) about 80\% of potential
candidates failed to fall within the catalog 2MFGC (see also Table\,1).
Apparently, when passing to the limit of the  2MASS survey, the galaxies look
more round and are cut off by the selection criterion. Note that for the
 RFGC galaxies 100\% completeness is achieved at $a_O=0\farcm85$ and drops
with the limiting diameter approaching $a_O=0\farcm6$.

\begin{table}[t]
{Table 1: {\it 2MFGC galaxies with different apparent magnitudes and
angular diameter}}
\begin{tabular}{|c|r|r|r|r|r|} \hline
&
\multicolumn{4}{c|}{2r, arcsec}&\\ \cline{2-5}
\multicolumn{1}{|c|}{J}&
\multicolumn{1}{|c|}{$<50$}&
\multicolumn{1}{|c|}{50--99}&
\multicolumn{1}{|c|}{100-149}&
\multicolumn{1}{|c|}{$\geq150$}&
\multicolumn{1}{|c|}{All} \\ \hline
$<$9   & 0     & 0    & 2  &  21 &   23  \\
9--10   &0      &2     &25  & 58  &  85  \\
10--11  &0      &112   &184 & 85  & 381  \\
11--12  &56     &918   &292 & 12  & 1278 \\
12--13  &1180   &2387  &84  & 0   & 3651 \\
13--14  &5171   &1767  &3   &0    &6941 \\
14--15  &4981   &212   &0   &0    &5193 \\
$\geq15$& 332   &1     & 0  & 0   & 333   \\

All     & 11720 &5399  &590 & 176 &17885 \\ \hline
H       &       &      &    &     &       \\    \hline
$<$9    &0     &0     &14   &61   &73       \\
9--10   & 0     &48    &134 &  82 &  264     \\
10--11  & 22    &668   &306 &  33 &  1029   \\
11--12  & 738   &2123  &123 &  0  &  2984  \\
12--13  &4267   &2091  &13  &  0  &  6371  \\
13--14  &5923   &474   &0   &  0  &  6397  \\
14--15  &739    &5     &0   &  0  &  744   \\
$\geq15$&  6    &0     &0   &  0  &  6  \\
All     &11695  &5409  &590 &  176&  17870 \\ \hline
K       &       &      &    &     &       \\  \hline
$<$9    &0      &2     &32  &  84 &  118  \\
9--10   &0      &156   &200 &  79 &  435  \\
10--11  &82     &1118  &279 &  13 &  1492 \\
11--12  &1766   &2500  &76  &  0  &  4342  \\
12--13  &6140   &1493  &3   &  0  &  7636  \\
13--14  &3704   &153   &0   &  0  &  3857  \\
14--15  &139    &0     &0   &  0  &  139   \\
$\geq15$ &1      &0     &0   &  0  &  1 \\
All     &11835  &5422  &590 &  176&  18020 \\ \hline
\end{tabular}
\end{table}

\begin{table}
{Table 2:{\it Distribution of 2MFGC galaxies according to angular diameters
for different axis ratios}}
\begin{tabular}{|c|r|r|r|r|r|} \hline
&
\multicolumn{4}{c|}{2r, arcsec}&\\ \cline{2-5}
\multicolumn{1}{|c|}{sab}&
\multicolumn{1}{|c|}{7--49}&
\multicolumn{1}{|c|}{50--99}&
\multicolumn{1}{|c|}{100-149}&
\multicolumn{1}{|c|}{$\geq150$}&
\multicolumn{1}{|c|}{All} \\ \hline
1--2.99& 199   & 170  & 20 &  10 &   399 \\
3--4.99 &11216  &4077  &359 & 109 &  15761\\
5--10   &417    &1175  &211 & 57  & 1860 \\   \hline
$a/b$   &       &      &    &     &       \\    \hline
1--2.99        & 4378  &421   &19  &  9  &  4827    \\
3--4.99        & 7394  &4291  &386 &  107&  12178  \\
5--10          & 60    &710   &185 &  60 &  1015 \\    \hline
All     &11832   &5422  &590 &  176&  18020\\  \hline
\end{tabular}
\end{table}

Another cause of omission of a ``really flat'' galaxy in the Catalog is
also possible. By our condition galaxies with $b/a=1$ were rejected. Thus,
the galaxies that look in the B band like Sc--Sd with a spherical high contrast
nucleus and a very faint disc component look in the  2MASS survey like compact
round objects and are discarded by the selection criterion.

In Fig.\,3 is present the distribution of 17885  2MFGC galaxies versus
their color $J-K$.
It looks quiet symmetric, with an average color
$<J-K> = 1\fm17$ and $\sigma=0\fm20$. This value is actually coincident
with the mean value of the color $<J-K> \simeq 1\fm2$, obtained for the
RFGC galaxies identified with XSC (Karachentsev et al.
2002). About 70\% of the 2MFGC galaxies have the color $J-K$ в in a narrow
interval, $0\fm9-1\fm3$.

The distribution of the  2MASS galaxies over the inverse axial ratio $a/b=1/sba$
(Fig.\,4) shows that with a sufficiently wide range of axial ratios the
greater part ($\geq$80\%) of our galaxies have $a/b\geq3$. The median value
$1/sab=4$, which is close to the value $med~(a/b)=4.1$ for the
RFGC galaxies identified in XSC.

In the interval of  $a/b$ from 4.5 to 10 an exponential relationship
$lg N = -0.42(a/b)+5.26 $ is satisfied. In Fig.\,4 there present nearly
round galaxies, although, according to the criterion of selection, objects with
$b/a=1$ were excluded. This effect takes place in adding up
$J, H$ and $K$ images of galaxies, when the elliptical isophotes describing
circumnuclear region are added together as averaged at different position
angles (PA) and produce practically round images.

Table\,2 gives the distribution of the numbers of the 2MFGC galaxies according
to angular dimensions depending on the fact which way of calculating the
axial ratios is adopted when selecting galaxies for the Catalog. With a ``soft''
criterion  $(b/a\leq0.34$) the number of nearly round galaxies that may be
entered into the Catalog is by a factor of 12 larger than with a ``hard''
criterion $sba\leq0.34$. As it was to be expected this effect is most
pronounced for the smallest $(2r\leq1\arcmin$) galaxies.

The distribution of the 2MFGC galaxies by near-infrared (NIR) position angles (PA)
 is shown in Fig.\,5. The mean number of galaxies per
interval is 1002 with the standard deviation of 31.6. The maxima and minima in
the distributions are insignificant. Only one value of 18 is beyond the
 $2\sigma$ range at the  PA from +40$\degr$ to $+50\degr$. Note that the position
angle is the least accurately determined parameter
presented in the Catalog. Comparison of the PA that we determined visually from
2MASS images with the
PA presented in XSC for a considerable share of galaxies shows a large (to
tens of degrees) deviations.
Rahman \& Shandarin
(2003) write about the ``twisting'' PA too. The position angles given in XSC
most likely refer to the circumnuclear region (bulge) than to galaxies as
a whole, as in RFGC.

\vspace*{-0.3cm}
\subsection{Some two-dimensional distributions}
It can be seen in Fig.\,6 that the average color $J-K$ for the 2MASS galaxies with
passing from bright to faint objects changes only slightly
(from 1$\fm$0 to 1$\fm$3), however the scatter in
$ J-K$  largely increases for $J > 13^m$. Nevertheless,
the average color of bright galaxies is bluer than that of faint ones. This can be
explained by the fact that, firstly, for bright galaxies the greater portion of
the disc component is involved in photometrying in comparison with faint
galaxies, and, secondly, faint galaxies are situated at larger distances,
where the K correction plays an important role.

 The axial ratio measured
in XSC changes slightly for both bright and faint
2MFGC galaxies (Fig.\,7). Note that our criterion of selection excluded the
brightest and round objects, which is indicated by the ``shortage'' of points
in the upper left corner of Fig.\,7.

The axial ratios measured for galaxies with different
$J-K$ colors change as slightly (Fig.\,8). Note the shortage of points among
the reddest and round galaxies.

Fig.\,9 shows the relationship of $J$ vs $\lg r$. Beginning with $J \leq13^m$,
it is seen that the galaxy sizes are systematically underestimated for the
brightest 2MASS objects. This is connected both with the nature of objects
selected for 2MFGC (late spirals have an extended blue disc component poorly
visible in the IR range) and, possibly, with characteristics of the digital
processing of extended objects.

In Figs.\,10--12  are exhibited two-dimensional distributions ``concentration
index'' IC vs $J$ magnitude, IC vs $(J-K)$ color and IC vs axial ratio,
respectively.

It is seen on all three diagrams that the mean index of concentration changes
slightly over a wide range of magnitudes and colors, having an average
value $<IC>=4.30$ with a standard deviation of $\sigma =0.93.$
Fig.\,10 shows deficiency of bright objects. Apparently, bright galaxies with
a large index of concentration failed to enter into the catalog
2MFGC for two reasons: 1) the selection criterion cut off objects with a
pronounced bulge (as required in selecting late spirals) and 2) the total
magnitude of thin galaxies might be underestimated in
photometrying.

The shortage of blue objects with ($J-K$) $<0\fm85$  at IC $>5$ (Fig.\,11) is
naturally explained by the absence of pronounced bulges in thin spirals of
late types.

Closer examination discloses that the index of concentration somewhat
decreases for galaxies with faint magnitudes,
$IC =-0.118J +5.886, \sigma_{regr} = 0.919,$
the correlation coefficient $r=-0.15$ and somewhat increases with ``reddening''
of a galaxy, $IC=0.651(J-K)+3.540, \sigma_{regr} =0.918$,
the correlation coefficient $r$ = 0.14.

It can be seen from Fig.\,12 that the mean value $IC=4.30$ remains unchanged
over the entire interval of $sba$ from 0.10 to 0.50. The main body of the
catalog 2MFGC is comprised of rather thin objects
($sba=0.27\pm0.04$).

\vspace*{-0.3cm}
\subsection{Distribution over the sky and NIR dipole }

We present in Fig.\,13 the distribution of
18020 galaxies of the catalog 2MFGC over the sky in supergalactic coordinates.
The gray strip isolates the region of the equator of the Galaxy
$(\mid b\mid <10\degr$). This region is seen to be sufficiently well filled
with galaxies.

In the lower right part of Fig.\,13 the positions of the Great Attractor (GA),
of the supercluster Pisces--Perseus (PP), Shapley  concentration of
 clusters (Sh),
and also
Bootes Void (BV) and Local Void (LV) are marked.  It follows clearly from
Fig.\,13 that the galaxies of the 2MFGC catalog outline a large-scale structure
at a great depth (for instance, radial velocities of clusters in the Shapley
concentration are $(14-20)\cdot 10^3$\,km/s). The same is also seen
at  nearer sections in distance,
up to $J \leq12^m$ (Fig.\,14 and 15).

As it is followed from Figs.\,13--15, the 2MFGC contains a large number of
galaxies distributed throughout the sky to a sufficient depth (the typical
redshift is $z\sim0.05$). This, as well as the photometric homogeneity of the
catalog 2MFGC, allows one to calculate the dipole moment in the distribution
of galaxies and make comparison with the results obtained
 from other catalogs (IRAS, RFGC).

%\vspace*{-0.3cm}
\begin{table*}[t]
\begin{center}
{Table 3: {\it NIR dipole parameters from the data of 2MFGC catalog }}
\begin{tabular}{|r|c|c|c|c|c|c|r|r|}  \hline
 K &  N   & X               &  Y               & Z               &$l\degr$& $b\degr$& $SGL\degr$& $SGB\degr$\\ \hline
10 & 554  & $-$0.023$\pm$0.022&  $-$0.137$\pm$0.023& 0.110$\pm$0.027 &260    & 38  &   122 & $-$36  \\
11 & 2070 & $-$0.049$\pm$0.012&  $-$0.054$\pm$0.012& 0.064$\pm$0.014 &227    & 41  &   90  & $-$43   \\
12 & 6423 & $-$0.049$\pm$0.007&  $-$0.030$\pm$0.007& 0.044$\pm$0.008 &211    & 37  &   73  & $-$43 \\
13 & 14078& $-$0.040$\pm$0.005&  $-$0.025$\pm$0.005& 0.029$\pm$0.005 &212    & 31  &   70  & $-$50  \\
14 & 17885& $-$0.041$\pm$0.004&  $-$0.016$\pm$0.004& 0.024$\pm$0.005 &201    & 28  &   56  & $-$47  \\
All& 18020& $-$0.043$\pm$0.004&  $-$0.014$\pm$0.004& 0.025$\pm$0.005 &198    & 28  &   53  & $-$45 \\ \hline
\end{tabular}
\end{center}
\end{table*}
Using the coordinates of the  2MFGC galaxies, we computed the location of
their centroid at different depths (different sections by apparent magnitude)
without making allowance for weighting galaxies by their brightness. The
results are listed in Table\,3. Here, in the appropriate columns are given:
$K$ --- the limiting magnitude to which the calculation of the dipole is
performed, N --- the number of galaxies in the subsample,
$X, Y$ and $Z$ --- the mean galactic Cartesian coordinates for the centroid of
galaxies and their errors, $l, b$,  $SGL$, $SGB$  --- the direction
of the galaxy centroid in galactic and supergalactic coordinates,
respectively. It is evident that with increasing depth (and rising
incompleteness) of the  2MFGC catalog the galactic longitude $l$ of the dipole
has a strong trend, while the galactic latitude changes slightly. Since the
catalog 2MFGC is assumed to be complete to $K=11^m$, then the computed positions
of the dipole for $K = 10^m$ and $K = 11^m$ are in fair agreement with the
data for the  IRAS dipole, $l=250\degr, b=38\degr$ (Lahav et al. 1988).

Maller et al. (2003) determined the position of the dipole as
$l=278\degr, b=+38\degr$ for objects brighter than $K=13\fm 5$ from the XSC
catalog (with taking account weighting by luminosity). As one can see from
our Table\,3, the dipole positions defined from the brightest ($K<10^m$)
2MFGC galaxies are in good agreement with the data of
Maller et al. (2003) too.

We present for comparison also the position of the MBR (Microwave Background
Radiation) dipole $l=268\degr, b=27\degr$, or $SGL =138\degr, SGB=-38\degr$
(Kogut et al. 1993).

Another group of results refers to the optical dipole. For the integrated
UGC+ESO  catalog (N = 23984, $a_{lim}=1\farcm03$) Lahav et al. (1998) give
the position of the sample centroid
 $l=261\degr, b=29\degr$.
For 4236 RFGC galaxies we obtained $l=273\degr, b=19\degr$  at $a_{lim}=
0\farcm6$ (Karachentsev et al. 1999).

The position of the  NIR  dipole found from truncated samples of the
catalog 2MFGC fits that of the MBR dipole, as well as the position of the
optical dipoles, within $\pm30\degr$.

\vspace*{-0.2cm}
\begin{acknowledgements}
We thank Dmitry Makarov for help in the design of Figs.\,13--15.

This paper makes use of data from the Two Micron All-Sky Survey, which
is a joint project of the University of Massachusetts and the Infrared
Processing and Analysis Center/California Institute of Technology, funded
by the National Aeronautics and Space Administration and the National
Science Foundation.

We have made use of the LEDA database (http://leda.univ-lyon1.fr) and
the NED database (http://nedwww.ipac.caltech.edu).

This research was partially supported by DFC-RFBR grant 02--02--04012.
\end{acknowledgements}

{}

\newpage

\begin{figure*}
\centerline{\psfig{figure=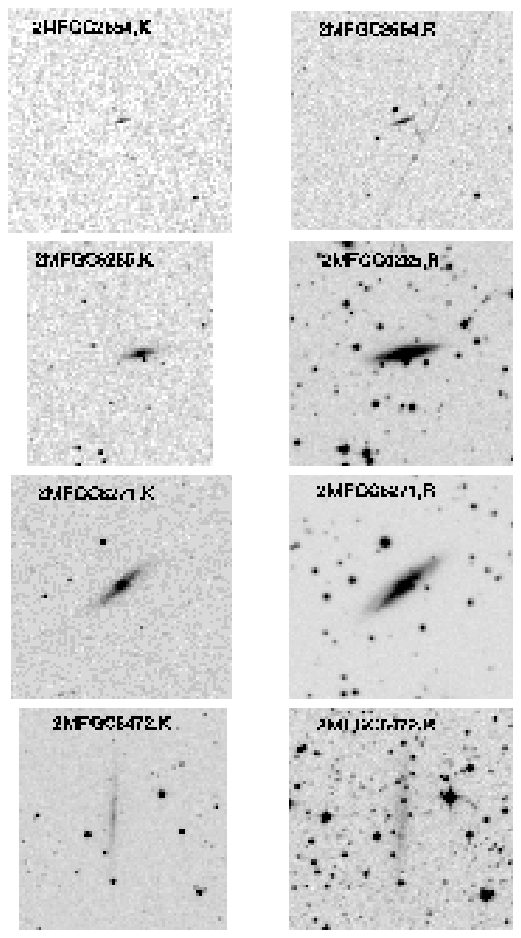,width=12cm}}
%\centerline{\psfig{figure=Fig1a.PS,bbllx=104pt,bblly=72pt,%
%bburx=496pt,bbury=741pt,clip=}}
\begin{center}
Fig1a: {\it Images of 2MFGC galaxies. On the left --- the view of a galaxy
in the  K band taken from 2MASS, on the right --- the view of the same
galaxy in the red band from DSS. The field size is $4\arcmin\times4\arcmin$.
For all the images North is top and East is left.}
\end{center}
\end{figure*}

\begin{figure*}
\centerline{\psfig{figure=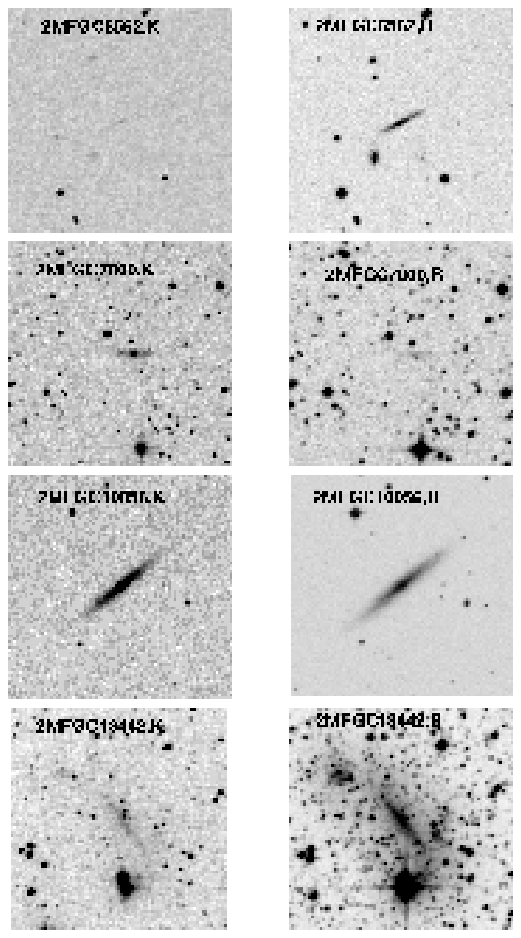,width=10cm}}
%\centerline{\psfig{figure=Fig1b.PS,bbllx=112pt,bblly=73pt,%
%bburx=500pt,bbury=745pt,clip=}}
\begin{center}
 Fig1b:{\it Images of 2MFGC galaxies (continuation of Fig.1a)}
\end{center}
\end{figure*}

\newpage
\setcounter{figure}{1}
\begin{figure*}
\centerline{\psfig{figure=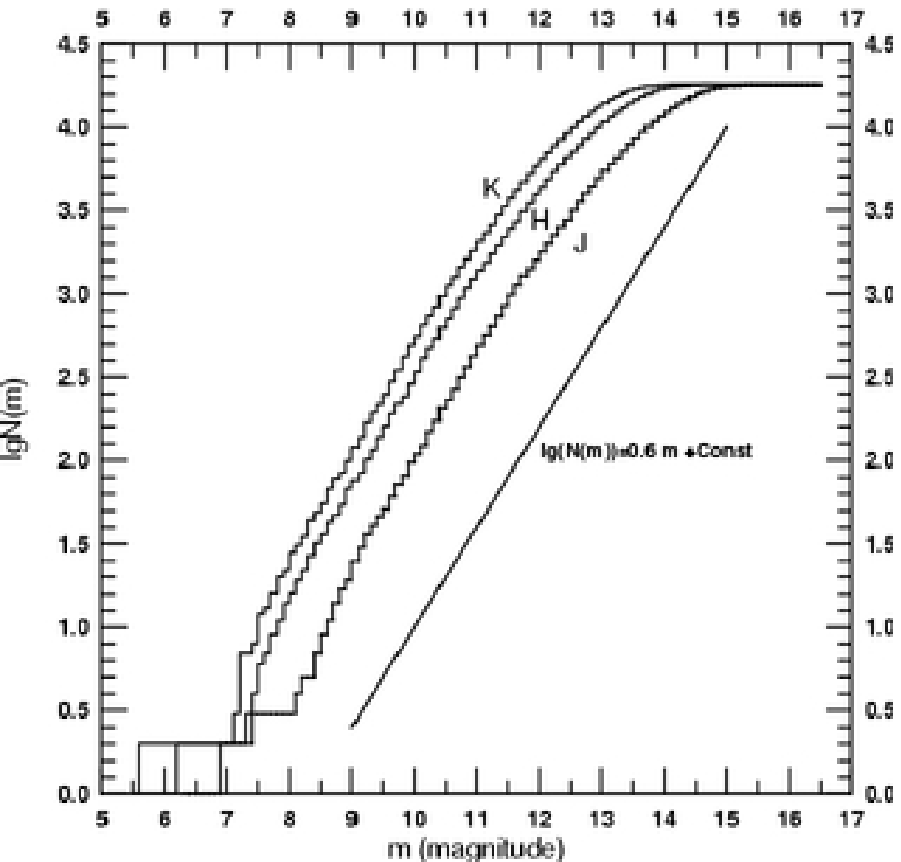,width=10cm}}
\caption
{The integral distribution of the 2MFGC galaxies on their NIR
magnitudes.}
\end{figure*}

\setcounter{figure}{2}
\begin{figure*}
\centerline{\psfig{figure=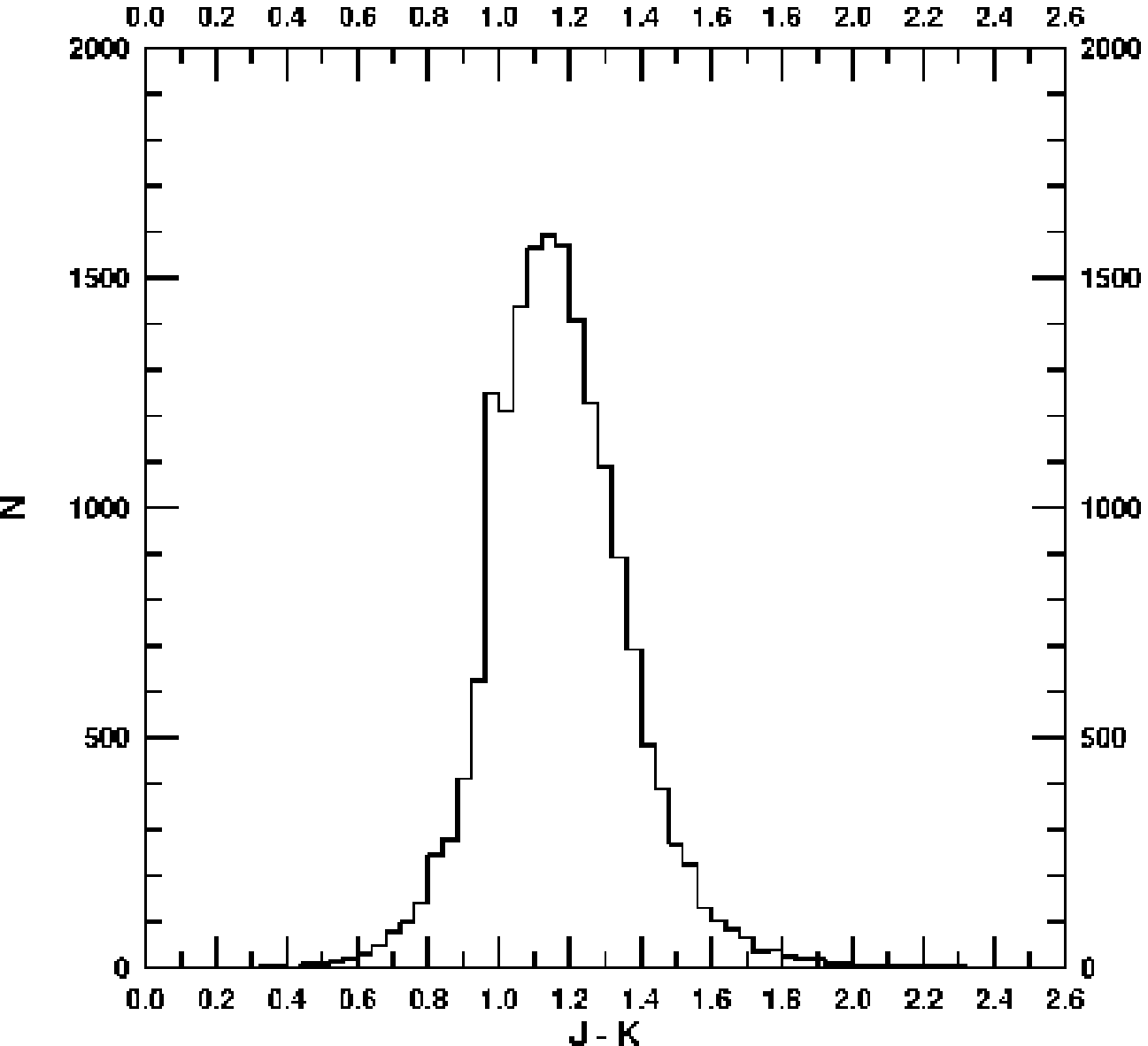,width=11cm,bbllx=20pt,bblly=23pt,%
bburx=408pt,bbury=384pt,clip=}}
\caption{
 The distribution of the 2MFGC galaxies on their J -- K colours.}
\end{figure*}

\newpage
\setcounter{figure}{3}
\begin{figure*}
\centerline{\psfig{figure=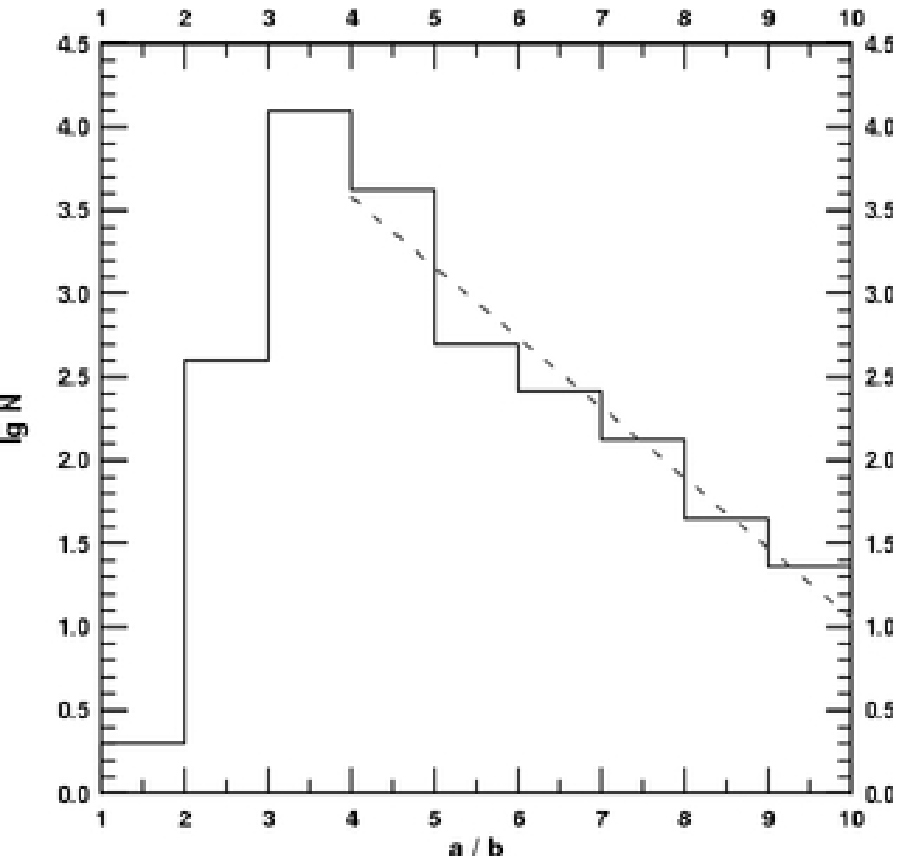,width=10cm}}
\caption{
The distribution of the 2MFGC galaxies on their axis ratio $a/b=1/sba$.
Dashed line marks the exponential law.}
\end{figure*}

\setcounter{figure}{4}
\begin{figure*}
\centerline{\psfig{figure=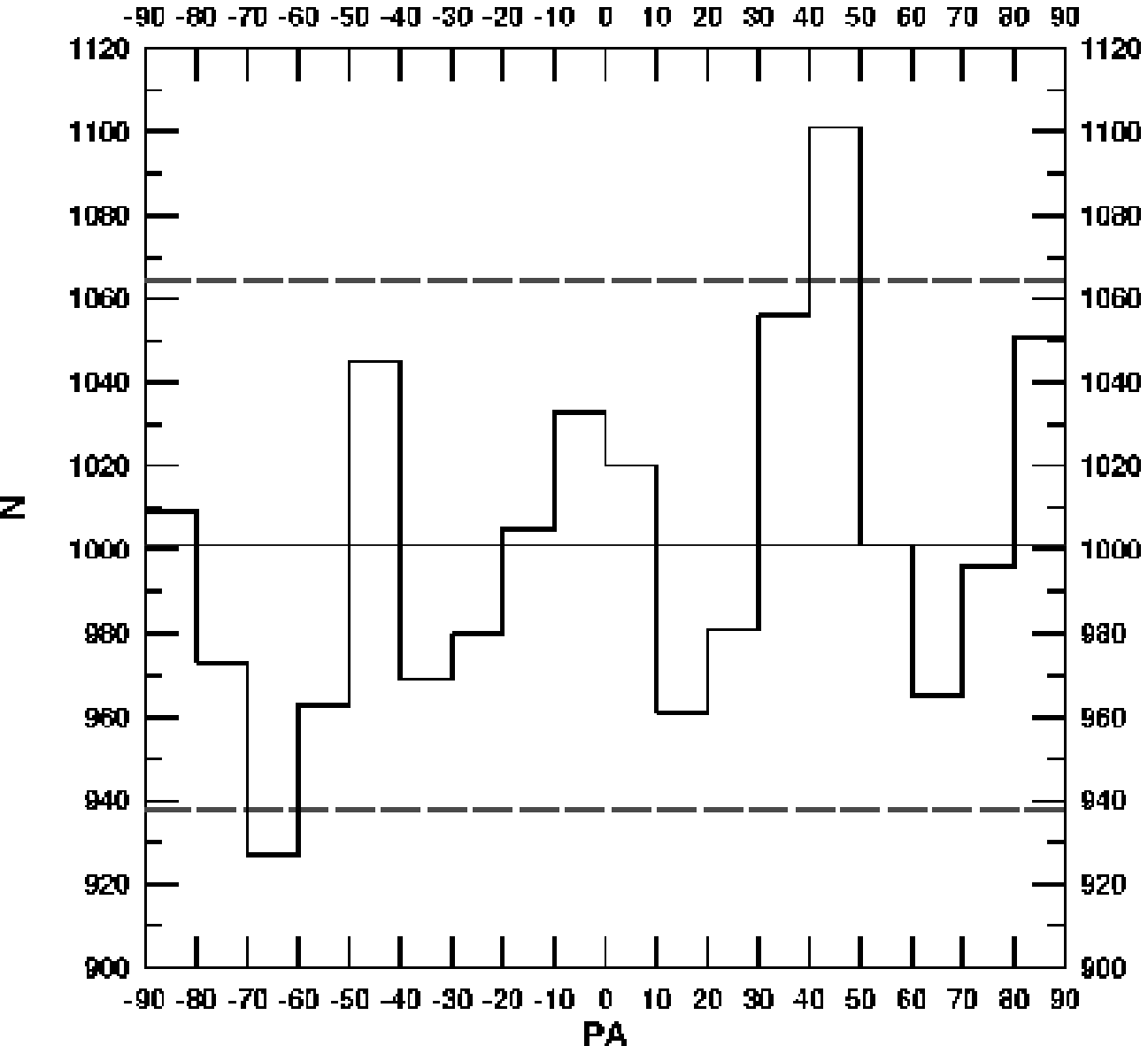,width=11cm,bbllx=19pt,bblly=25pt,%
bburx=408pt,bbury=380pt,clip=}}
\caption{
 The distribution of the 2MFGC galaxies on their position angles.
The $\pm2\sigma$ level is marked by dashed lines, and the mean one --- by
solid line.}
\end{figure*}

\newpage
\setcounter{figure}{5}
\begin{figure*}
\centerline{\psfig{figure=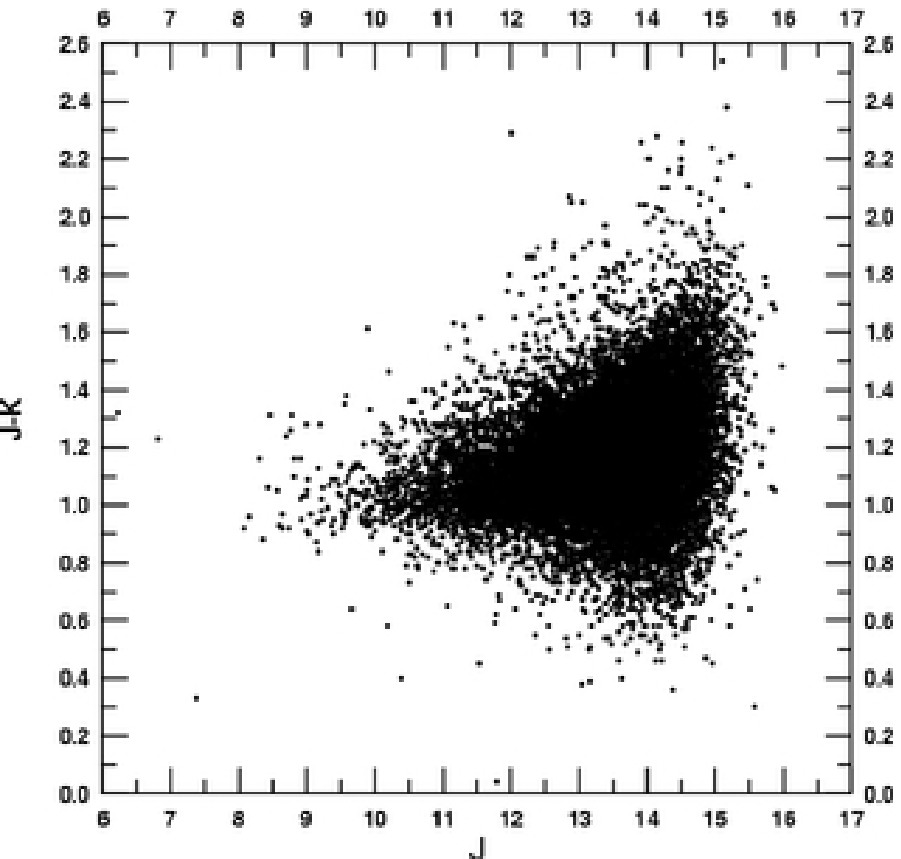,width=10cm}}
\caption{
 The (J -- K) colour on J magnitude relationship for 2MFGC galaxies.}
\end{figure*}

\setcounter{figure}{6}
\begin{figure*}
\centerline{\psfig{figure=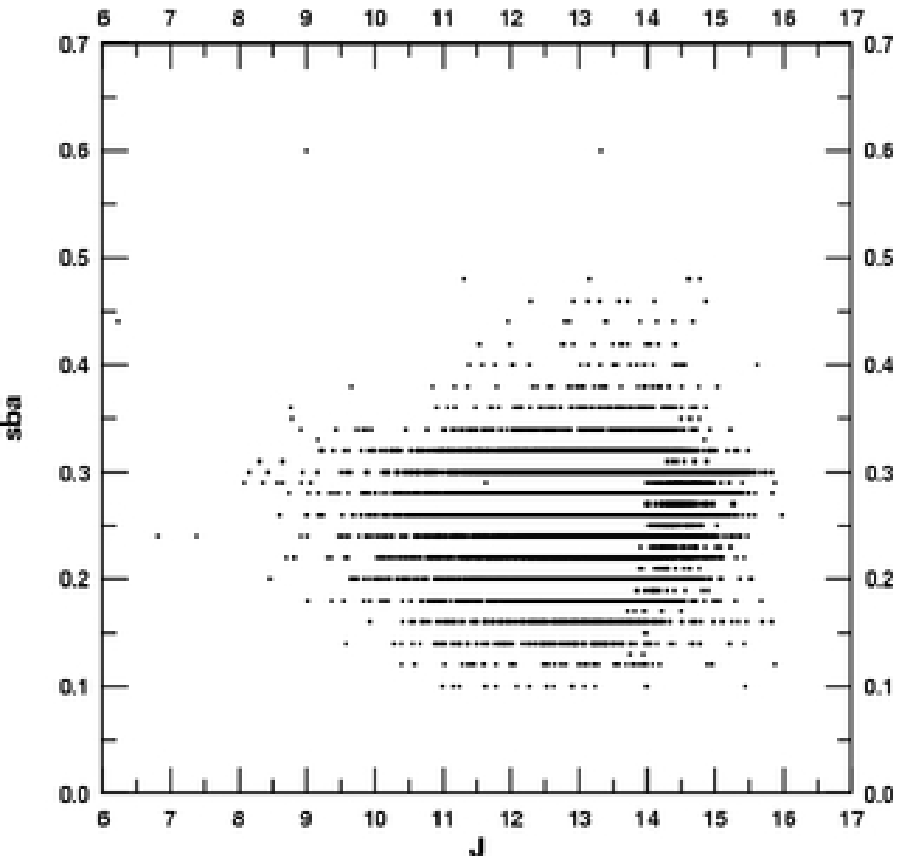,width=10cm}}
\caption{
 The relationship between the ``super'' coadd axis ratios and the J magnitudes
for 2MFGC galaxies.}
\end{figure*}

\newpage
\setcounter{figure}{7}
\begin{figure*}
\centerline{\psfig{figure=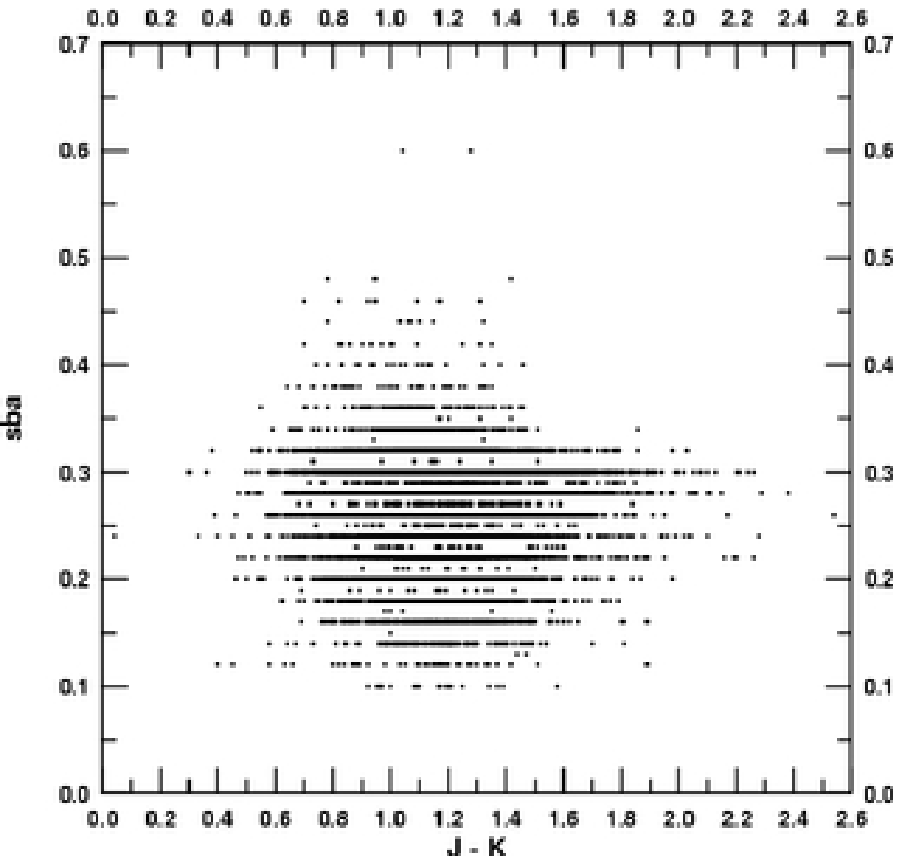,width=10cm}}
\caption{
 The dependence of the ``super'' coadd axis ratios on the (J -- K)
colours for  2MFGC galaxies.}
\end{figure*}

\setcounter{figure}{8}
\begin{figure*}
\centerline{\psfig{figure=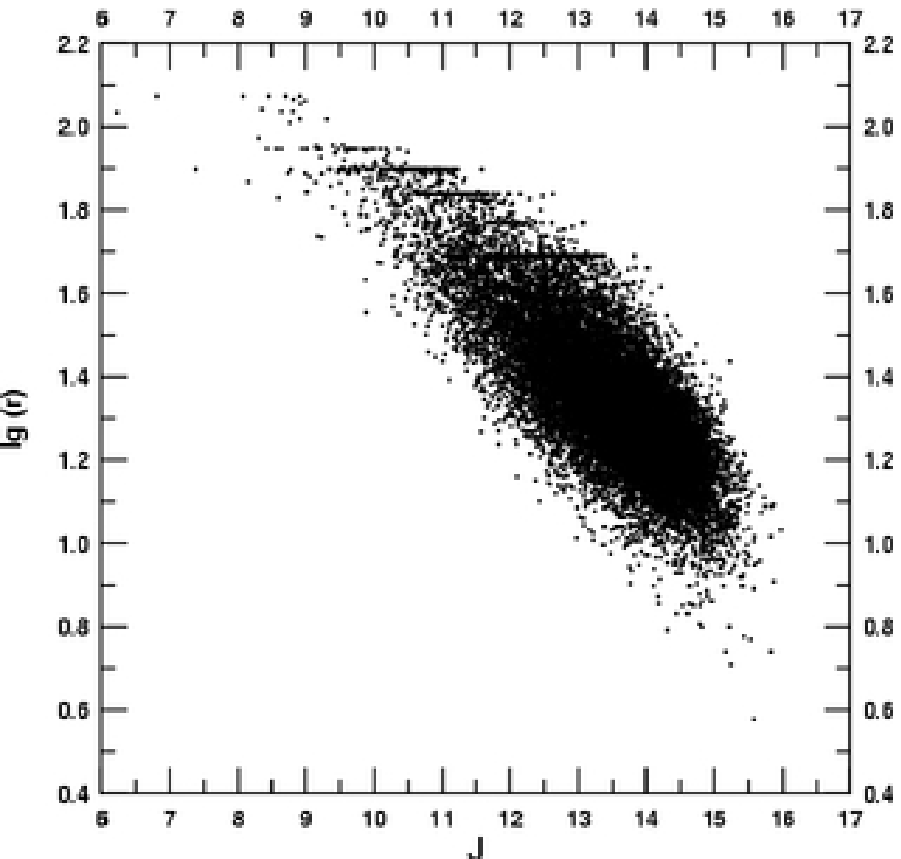,width=10cm}}
\caption{
 The relation between $\lg(r)$ and J magnitude for 2MFGC galaxies.}
\end{figure*}

\newpage
\setcounter{figure}{9}
\begin{figure*}
\centerline{\psfig{figure=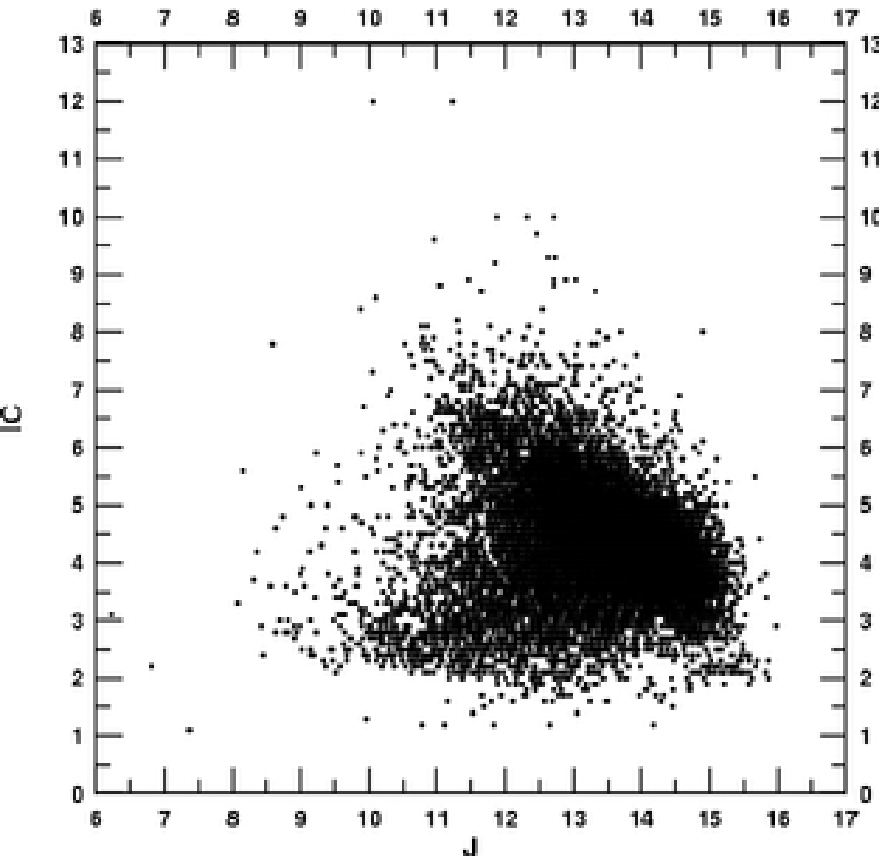,width=10cm}}
\caption{
 The relation between the concentration index and J magnitude
for 17880 2MFGC galaxies.}
\end{figure*}

\setcounter{figure}{10}
\begin{figure*}
\centerline{\psfig{figure=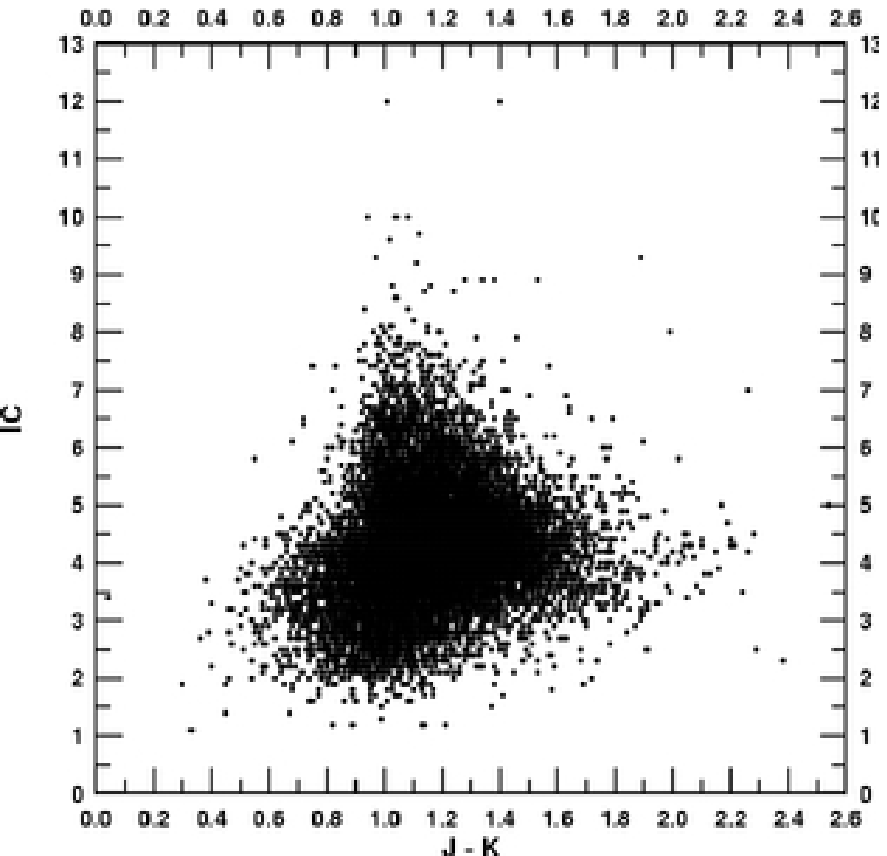,width=10cm}}
\caption{
 The relation between the concentration index and J -- K color
for 17880 2MFGC galaxies.}
\end{figure*}

\newpage
\setcounter{figure}{11}
\begin{figure*}
\centerline{\psfig{figure=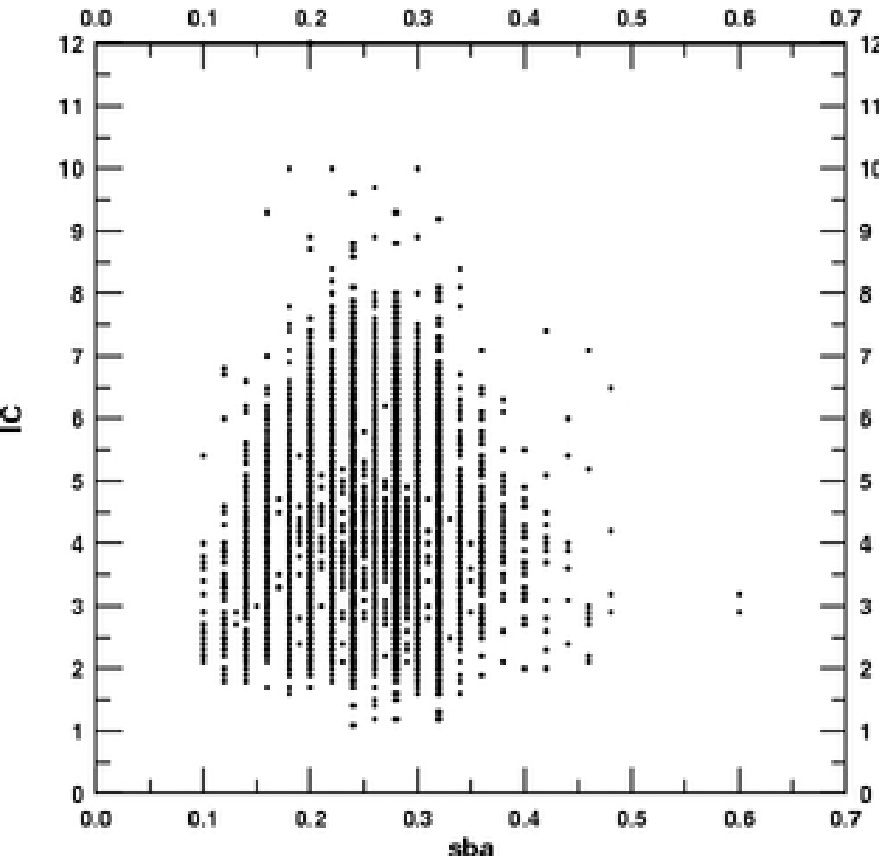,width=14cm}}
\caption{
 The relation between the concentration index and ``super'' axis
ratio for 18015 2MFGC galaxies.}
\end{figure*}

\setcounter{figure}{12}
\begin{figure*}
\centerline{\psfig{figure=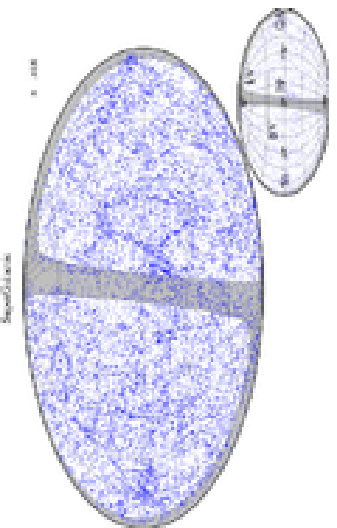,height=12cm}}
\caption{
 The distribution of the 2MFGC galaxies over the sky in supergalactic
coordinates. The region with $\mid b\mid=10\degr$ is marked by grey colour.}
\end{figure*}
\newpage
%\end{document}

\setcounter{figure}{13}
\begin{figure*}
\centerline{\psfig{figure=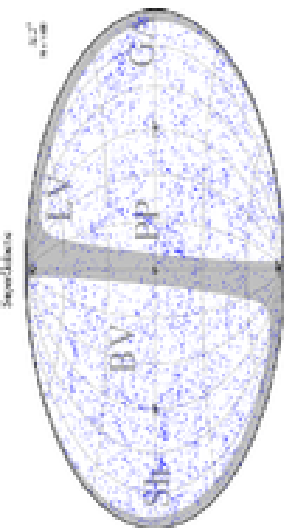,height=12cm}}
\caption{
 The distribution of the 2MFGC galaxies with J $\leq13$ over
the sky in supergalactic coordinates. The positions of well-known galaxy
concentrations and voids are marked.}
\end{figure*}
\newpage
\setcounter{figure}{14}

\setcounter{figure}{14}
\begin{figure*}
\centerline{\psfig{figure=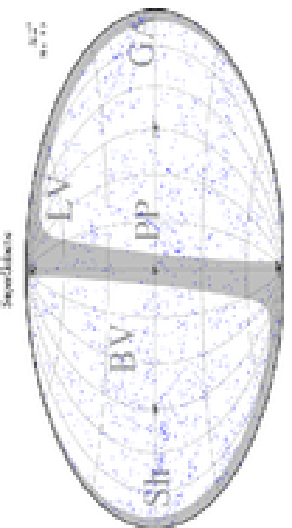,height=12cm}}
\caption{
 The distribution of the 2MFGC galaxies with J $\leq12$ over
the sky in supergalactic coordinates. The designations of the galaxy
concentrations and voids are the same as in Fig.13 and 14.}
\end{figure*}


\vspace*{-0.5cm}
\begin{thebibliography}{}
\bibitem {}Bizyaev D. \& Mitronova S., 2002, Astron.Astrophys., {\bf 389}, 795
\bibitem {} Curti R.M., Skrutskie M.F., 1998, Two Micron All-Sky Syrvey Status Report,
   BAAS, {\bf 30}, 1374
\bibitem{} Jarrett T.H., 2000, Publ. Astr. Soc. Pacific, {\bf 112}, 1008
\bibitem{} Jarrett T.H., Chester T., Cutri R. et al., 2000, Astron.J.,
{\bf 119}, 2498
\bibitem{} Jarrett T.H., Chester T., Cutri R., Schneider S., Huchra J.,
 2003, Astron.J., {\bf 125}, 525
\bibitem{} Karachentsev I.D., 1989, Astron.J., {\bf 97}, 1566
\bibitem{} Karachentsev I.D., Karachentseva V.E., Parnovsky S.L., 1993,
Astron. Nachr., {\bf 314}, 97 (FGC)
\bibitem{} Karachentsev I.D., Karachentseva V.E., Kudrya Yu.N. et al.,
1997, Pis'ma  Astron. Zh., {\bf 23}, 652
\bibitem{} Karachentsev I.D., Karachentseva V.E., Kudrya Yu.N. et al.,
1999, Bull. Spec. Astrophys. Obs., {\bf 47}, 5 (RFGC)
\bibitem{} Karachentsev I.D., Mitronova S.N., Karachentseva V.E. et al., 2002,
Astron. \& Astrophys. {\bf 396}, 431
\bibitem{} Karachentsev I.D., Karachentseva V.E., Kudrya Yu.N. et al.,
2000, Astron. Zh., {\bf 77}, 175
\bibitem{} Kogut A., Lineweaver C., Smoot G.F. et al., 1993, Astrophys. J.,
{\bf 419}, 1
\bibitem{} Kudrya Yu.N., Karachentsev I.D., Karachentseva V.E. et al.,
1997a, Pis'ma Astron. Zh., {\bf 23}, 15
\bibitem{} Kudrya Yu.N., Karachentsev I.D., Karachentseva V.E. et al.,
1997b, Pis'ma Astron. Zh., {\bf 23}, 730
\bibitem{} Kudrya Yu.N., Karachentsev I.D., Karachentseva V.E. et al.,
2003, Astron. \& Astrophys., {\bf 407}, 889
\bibitem{} Lahav O., Rowan-Robinson M., Lynden-Bell D., 1988, \mnras,
{\bf 234}, 667
\bibitem {} Lauberts A., 1982, The ESO/Uppsala Survey of the ESO(B) Atlas,
ESO, Munich
\bibitem {} Maller A.N., McIntosh D.N., Katz N. et al., 2003, astro-ph/0303592
\bibitem{} Nikolaev S., Weinberg M.D., Skrutskie M.F. et al., 2000,
Astron. J., {\bf 120}, 3340
\bibitem{} Nilson P., 1973, Uppsala General Catalogue of Galaxies,
Uppsala Astron. Obs. Ann., Bd.6 (UGC)
\bibitem{} Raman \& Shandarin S., 2003, astro-ph/0310242
\bibitem{} Skrutskie M.F., Schneider S.E., Stiening R. et al., 1997,
in: ``The Impact of Large Scale Near--IR Sky Surveys, ed.
F. Garzon et al., Netherlands: Kluwer, ASSL, {\bf 210}, 25
\bibitem{} Vaucouleurs G., de, Vaucouleurs A. de, Corwin H.C. et al.,
Third Reference Catalogue of Bright Galaxies, New York, Springer, 1991,
{\bf 1-3}
\bibitem{} Zasov A.V., Bizyaev D.V., Makarov D.I. et al., 2002, Pis'ma
Astron. Zh., {\bf 28}, 1
\end{thebibliography}
\end{document}